\def\ra{\rightarrow}
\documentstyle [aps,preprint,pre] {revtex}
\input epsf
\begin {document}
\draft
\title {Breakdown of Simple Scaling in
Abelian Sandpile Models in One Dimension}
\author {Agha Afsar Ali and Deepak Dhar}
\address {Theoretical Physics Group \\
          Tata Institute of Fundamental Research, \\
          Homi Bhabha Road, Bombay 400 005, India.}
\date{\today}
\maketitle
\begin {abstract}
We study the abelian sandpile model on decorated one dimensional
chains. We determine the structure and the asymptotic form of
distribution of avalanche-sizes in these models, and show that
these differ qualitatively from the behavior on a simple linear
chain. We find that the probability distribution of the total number of
topplings $s$ on a finite system of size $L$ is not described by a
simple finite size scaling form, but by a linear
combination of two simple scaling forms $Prob_L(s) = {1 \over L}f_1({s
\over L})+{1 \over L^2}f_2({s \over L^2})$, for large $L$, where $f_1$
and $f_2$ are some scaling functions of one argument.
\end{abstract}

\narrowtext

In recent years there has been a lot of interest in the systems
showing self-organized criticality (SOC) \cite{btwo,dmo,po}. However
the precise conditions under which the steady state of a driven system
shows critical (long range) correlations are not well understood for
non-conservative systems \cite{ofc,ds,mt}. In the case of systems with
conservation laws \cite{hk,gls}, for example in sandpile models with
local conservation of sand, it is easily shown that the average
number of topplings in an avalanche diverges as a power of the system
size \cite{do,k}. This, however, is not sufficient to ensure
criticality, if criticality is defined as the existence of power law
tails in the distribution of avalanche sizes \cite {f1}.

Lacking a general theory, most studies of SOC depend upon numerical
simulations for evidence of criticality. To incorporate the effect of
finite size cut-offs, it is usual
to fit data to a finite-size scaling form in which the critical
exponents of the infinite system appear as
parameters. However, on the basis of extensive numerical studies of
one dimensional sandpile automata, Kadanoff and coworkers
\cite{knwz,cfkkp} have argued that there is more than one
characteristic length scale in many of these models. Consequently, a
simple finite size scaling does not describe the statistics of
avalanches, and a more general `multifractal' scaling form seems
necessary.

   As the finite-size scaling assumption based on a single scaling
form is widely used in the studies of SOC, it seems desirable to test
it in a simple analytically tractable model. This we do in this Rapid
Communication for a class of one dimensional Abelian Sandpile Models
(ASM).  We find that for large $L$, the distribution function of
duration $t$ of an avalanche, and the number of distinct sites toppled
$s_d$ in our model do have a simple scaling
form. However, the distribution function of total number of topplings
$s$, and of the maximum number of topplings $n_c$ at any one site does
{\it not} have a simple scaling form, but a more complicated linear
combination of two simple scaling forms (LC2SSF)
\begin{equation}
Prob_L(X) = L^{-\beta_1} f_1(XL^{-\nu_1}) + L^{-\beta_2} f_2
(XL^{-\nu_2}) ~~~,
\label{eq1}
\end{equation}
for large $L$, where $\beta_1 = \nu_1 = 0$ and $\beta_2 = \nu_2 = 1$
for $X=n_c$ and $\beta_1 = \nu_1 = 1$ and $\beta_2 = \nu_2 = 2$ for
$X=s$, and $f_1$ and $f_2$ are scaling functions, different for $X=s$
and $X=n_c$. We also find that this behaviour is quite robust and does
not depend on the choice of the unit cell, but in general the function
$f_1$ and $f_2$ are not universal.

   The ASM on a simple linear chain has been studied earlier by Bak
{\it et al} \cite{btwt}, and in more detail by Ruelle and Sen
\cite{rs}. We consider ASM on one dimensional
chains formed by joining a single type of unit cells (see
Fig. 1). Such decorated chains are the simplest generalization of the
linear chain. We have studied two cases in detail. Case A is a chain
of alternating double and single bonds. Case B is a chain of diamonds
joined together by single bonds.  We solve these models exactly in the
large $L$ limit, and find that the avalanche distribution function
shows a nontrivial behaviour, different from that of the simple linear
chain (case C). In fact the behavior of the ASM in case C is not
typical of one-dimensional ASM's.

 The model is defined as follows: A site on the chain is denoted by a
pair of indices $(i,j)$, where $i=1$ to $L$ labels the unit cell and
$j$ numbers a site within the unit cell. In case A $j$ ranges from 1
to 2, and in case B, from 1 to 4. At each site $(i,j)$ there is an integer
variable $h_{ij}$, called height of the sandpile at that site. A
particle is added at a randomly selected site. If the height
$h_{ij}$ is greater than a preassigned threshold height
$h_{ij}^c$ at that site it topples, and loses one particle to each
of its neighbours.  We choose $h^c_{ij}$ to be independent of $i$ and
equal to the coordination number of site of type $j$.  A toppling at a
boundary site causes a loss of one particle from the system. The
process of toppling continues until there are no unstable sites. After
the system is stabilized a new particle is added.

The critical steady state is easy to characterize using the general
theory of ASM's \cite{do}. In the steady state all recurrent
configurations occur with equal probability. The set of recurrent
configurations is characterized by the burning algorithm (see
\cite{dmo}, also
\cite{s}). In the burning algorithm, the sites can be burnt in any
order. We choose the convention that the burning starts from the left
boundary and continues rightward as long as possible.  The unit cell
where the rightward burning stops will be called the break point.
Afterwards, the burning is allowed to proceed leftwards from the right
boundary. It is easy to see that in a recurrent configuration of model
A, the allowed values of $(h_{i1},h_{i2})$, for $i$ on the left of
the break point, are $(3,3)$ and $(3,2)$. For $i$ on the right of the
break point these are $(3,3)$ and $(2,3)$ and at the break point these
are $(2,3)$, $(3,1)$ and $(1,3)$. Since each doublet other than the
break point has only two possible configurations, the entropy per
site, defined as the logarithm of the total number of recurrent
configurations divided by the number of sites, is finite and equal
$\ln(2) / 2$ in the large $L$ limit. For the simple linear chain,
the entropy per site in the SOC state is zero. This fact is
responsible for its non-generic behavior.

 To the left (right) of the break point the left (right) site of a
doublet always has height 3, and the probability of right (left) site
of a doublet having height 2 and 3 is $1 \over 2$ each. The break
point can occur at any of the L doublets with equal
probability. Averaging over the position of the break point, this
implies that the probabilities of the left site of $i\,th$ doublet
having height 2 and 3 are $i/ (2L)$ and $1-i/ (2L)$
respectively. Similarly the probabilities of the right site of a
doublet having height 2 and 3 are ${1\over2}(1-i/L)$ and
${1\over2}(1+i/L)$ respectively. Thus the average height profile in
the SOC state varies linearly with $i$ in case A, and the SOC state is
{\it not translationally invariant even far away from the
boundaries}. This feature is not present in case C.

Now we describe the spread of the avalanche in model A, which again
differs qualitatively from case C. Without loss of generality we may
assume that the point where the particle is added to be called the
source site, is to the left of the break point. Then clearly, if the
configuration of the doublet left to it is $(3,2)$, the avalanche does not
spread to the left and propagates a distance of order $L$ upto the
break point on the right. Each site affected by the avalanche topples
only once, and the total number of topplings in an avalanche is of
order $L$.
Such an avalanche is said to be of type
I. The probability of such avalanches is $1\over 4$. One can easily
check that otherwise the avalanche propagates a distance of order $L$
on both sides of the source point. In such avalanches $n_c$ is of
order $L$, and the total number of toppling in an avalanche is of
order $L^2$.
Such an avalanche is said to be of
type II (see Fig. 2). As the probability that the addition of particle
will cause an avalanche is $3 \over 4$, the fractional number of
avalanches of type I is $1\over 3$.

The probability distributions of total number of toppling $s$, total
number of distinct sites toppled $s_d$, duration $t$, and the number
of times the source site topples $n_c$, for type I avalanche can be
calculated easily. It is convenient to work with the scaled variable
$\alpha \equiv {i \over L}$ and $\beta \equiv {j \over L}$, such that
$\alpha,~\beta\in [0,1]$, where $i$ and $j$ are the position of source
point and break point on the chain respectively. It can be easily
verified that for type I avalanche $s$, $s_d$ and $t = 2(\beta -
\alpha)L$. Thus the probability distribution of $s/L$, $s_d/L$ and
$t/L$ for given $\alpha$ and $\beta$ for type I has a delta function
at $2(\beta -\alpha)$. Using the fact that $\alpha$ and $\beta$ are
independent random variables uniformly distributed between 0 and 1,
averaging over $\alpha$ and $\beta$ we find for type I avalanches
\begin{equation}
Prob_L(X | \hbox{type I}) = [1 - X/( 2L)]/L ~,~~~\hbox{for}~~~X\leq 2L~~,
\label{eqn2}
\end{equation}
where $X=s,s_d,t$. In type I avalanches any site topples at most once
so $n_c = 1$.

The type II avalanches show a much complicated and interesting
structure. The avalanche fronts, i.e. the left and right boundary sites
of the active region at any time, do not move uniformly in time, the
spreading rate depends on the local height configuration. However, for
distances $>> 1$, one can define an average velocity. The analysis of
these avalanches become easy using the decomposition of avalanches
into waves of toppling proposed by Ivashkevich {\it et al}
\cite{iv}. In each wave of toppling the source site topples only once
and all other sites topple until they are stable.  Waves of toppling
propagate in exactly the same way as the burning front in burning
algorithm.  Thus a unit cell which cannot be fully burnt from the left
(right) side stops a wave propagating towards left (right), and is
modified so that next wave may cross it.  We refer to such
configurations as left (right) stoppers. The stoppers slow down the
spreading of avalanches. Obviously the first wave propagates upto the
break point with a velocity 1 site per
time-step. To calculate the velocity towards left from the source
point we note that (a) doublet of type $(3,3)$ is crossed in 2 time
steps and (b) doublet of type (3,2) is crossed in 4 time steps because
it stops the first wave approaching to it and it is crossed in 2 time
steps by the next wave which follows after 2 time steps of the
previous wave. Thus the average time taken by the avalanche front to
cross a doublet is 3, which implies the average velocity is $2\over 3$
sites per time-step.  Similarly one can show that if the avalanche
crosses the break point on the right it will advance with an average
velocity which is also $2\over 3$ sites per time steps. The velocity
with which avalanche front recedes backwards after it has hit the
boundary is $2\over 3$. Details will be presented elsewhere.

Since the avalanche front moves with an average velocity, it forms a
polygon in the space time history of the avalanche (see Fig. 2)
\cite{f2}.
The number of sides in the polygon
depend on the position of the source point $\alpha$ and break point
$\beta$ and on whether the break point is crossed by the avalanche or
not.  For example if $\beta >\alpha > 5\beta/6$ and the break point is
not crossed then the polygon has only four edges. If $1-6\alpha
>\beta > \alpha$, and the break point is crossed then the polygon has
6 edges.  There are seven possible cases of polygons which need to be
analysed separately. Quantities like $s_d$, $t$ and $n_c$ which
are proportional to
the linear size of the polygon scale as $L$, and $s$ which goes as
area of the polygon scale as $L^2$.  The expressions of scaled
variables $s/L^2$, $s_d/L$, $t/L$ and $n_c/L$ can be easily evaluated
in terms of $\alpha$ and $\beta$ for each case.  The probability
distribution functions for given $\alpha$ and $\beta$ is a sum of two
delta functions corresponding to the cases whether the break point is
crossed by the avalanche or not.  Averaging over $\alpha$ and $\beta$
we find
\begin{equation}
Prob_L\,(q|\hbox{type II}) = \sum_{i=1,2}\int_0^1\int_0^1\,d\alpha \,
d\beta \,\, C_i\delta
\left ( q - q_i (\alpha ,\beta)\right )
\label{eqn4}
\end{equation}
where $q$ is the generic notation for $s/L^2$, $s_d/L$, $t/L$ and
$n_c/L$, $C_1=1/3$ is the probability that the a type II avalanche
crosses the break point and $C_2=2/3$ is the probability that it does
not cross the break point, and $q_1$ and $q_2$ denote expressions of
$q$ in terms of $\alpha$ and $\beta$ in the two cases. The full
explicit expression is quite complicated and will be presented
elsewhere.

However, some of the important features of the distribution function
can be understood by simple arguments. Since $s_d$ is the extension of
the polygon along horizontal axis, $s_d/L$ is a linear function of
$\alpha$ and $\beta$ in each of the seven cases. Hence the probability
distribution of $s_d/L$ is a piece wise linear function. The same argument
works for $t$ and $n_c$ also. The total number of topplings $s$ is
proportional to the area of the polygon. Therefore, $s/L^2$ is a
quadratic function of $\alpha$ and $\beta$. The probability
distribution in this case is quite complicated and diverges as
$(s/ L^2)^{-1/2}$ for small $(s/ L^2)$.

Summing over the contribution coming from type I avalanches (equation
(\ref{eqn2})) and type II avalanches (equation (\ref{eqn4})) we obtain
the full probability distributions. Since $n_c$ and $s$ scale
differently for type I and type II avalanches the distributions of
these quantities have the form given in equation (\ref{eq1}). Other
quantities like $s_d$ and $t$ scale as $L$ for both types of
avalanches. Therefore, the distribution of $s_d$ and $t$ have a simple
scaling form.

The treatment is easily extended to other types of unit cells
also. For example in case B the unit cell is a diamond. In this case
also, an avalanche always spreads upto the break point. The spread of
avalanches to the other side will be either of order $L$ or of order
1. Thus again, there are two types of avalanches. A detailed
calculation shows that these occur with relative frequencies $5:8$ on
the average.  While the velocities of avalanche front are different in
this case, the probability distribution functions for both type I and
type II avalanches have the same qualitative features irrespective of the
velocities.  For type I avalanches, $t \sim s_d
\sim (\beta-\alpha)L$ to order $L$. Thus probability distribution of
$s_d$ and $t$ have same linear form as in model A, while the slope
depends on the velocities. The variable $n_c$ has the probability
distribution $Prob(n_c) \sim 2^{-n_c}$. As $s \sim n_c(\beta
-\alpha)L$, this implies that the scaling function $f_1$ in Eq.
(\ref{eq1}), is a piecewise linear function with many segments.
For type II avalanches the space time history of active sites forms a
polygon exactly like in model A, except that the slope of edges of the
polygon depend on the velocities. Therefore the probability
distributions have same qualitative behaviour as in model A. However
the exact form of functions $f_1$ and $f_2$ are not same in case A and
B, and these functions are {\it not universal}. In case C, there are
no avalanches of type I, and the simple scaling ansatz works \cite{rs}.

In the multifractal approach one defines the function $f(\alpha)$ by
the relation that an avalanche of size $X = L^\alpha$ occurs with a
probability which scales as $L^{f(\alpha)}$, for large $L$.  The
exponent $f(\alpha)$ defined as
$\lim_{L\ra\infty}\log(Prob_L(X)/\log(L))$ is a continuous function of
the $\alpha$.  For our abelian model it is easy to see from Eq.
(\ref{eq1}) that $f(\alpha)$ is a monotonically decreasing piecewise
linear function for $X=s$ (see Fig. 3).  We have also shown results of
a computer simulation of the model for $L=100$ for $2 \times 10^5$
avalanches. Also shown is the theoretical curve using the Eq.
(\ref{eq1}) for $L = 100$ (dotted line) and $L= \infty$ (solid line).
Clearly
there is a very good agreement with simulation data. We note that The
$f$ versus $\alpha$ curve is quite similar to that obtained in
\cite{cfkkp} and that approach to $L\ra \infty$ limit is quite slow.

As the LC2SSF involves only a finite number of unknown
parameters, its use when simple scaling fails is preferable over the
more general multifractal form. We also note that we find the
breakdown of simple scaling without appearance of two different length
scales in our model.

Similar behaviour may be expected in other one dimensional models. For
example, for ASM on $L\times M$ cylinder $L>>M>>1$, we expect three types
of avalanches: type I and II, and finite avalanches of size $<M$,
which do not ring the cylinder, and are two dimensional in character.
This shows that a LC3SSF would describe this situation. It remains to
be seen whether this behavior survives in higher dimensions or it is
specific to one dimensional models.

To summarize, we have determined an exact asymptotic finite size
scaling behaviour of the distribution of avalanche sizes in the
abelian sandpile model on a class of decorated one dimensional
chains. We find that in these models the SOC state is not
translationally invariant, and the probability distribution of $s$ and
$n_c$ unlike the simple linear chain is described by a linear
combination of two simple scaling forms, and not by simple scaling
form.

We thank M. Barma and G. Menon for a critical reading of the
manuscript.

\newpage

{\bf Captions}

Figure 1: The one dimensional chains formed by joining (A) doublets,
(B) diamonds, (C) single sites.

Figure 2: The evolution of (a) type I avalanche, (b) type II
avalanche in model A. The filled rectangle denotes a toppling event.

Figure 3: The $\log-\log$ plot of $Prob(s)$ vs $s$.
The solid line shows the exact asymptotic behaviour for
$L \rightarrow \infty$, and the dotted line shows the
theoretical curve for $L = 100$.


\begin{thebibliography}{99}
\bibitem{f2} We ignore here the deviation from straightness in the
sides of the ``polygons'', as these are of order $\sqrt{l}$, where $l$
is the length of the side.
\end{thebibliography}
\end {document}
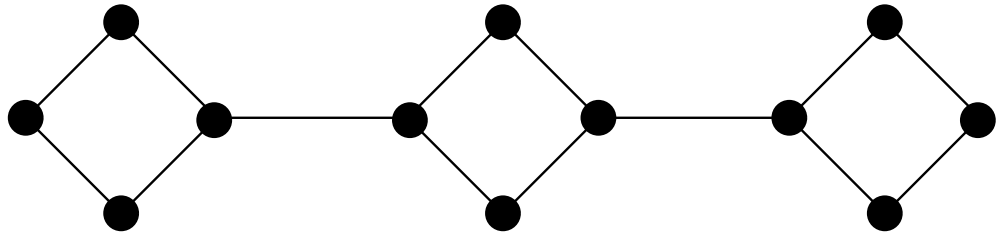
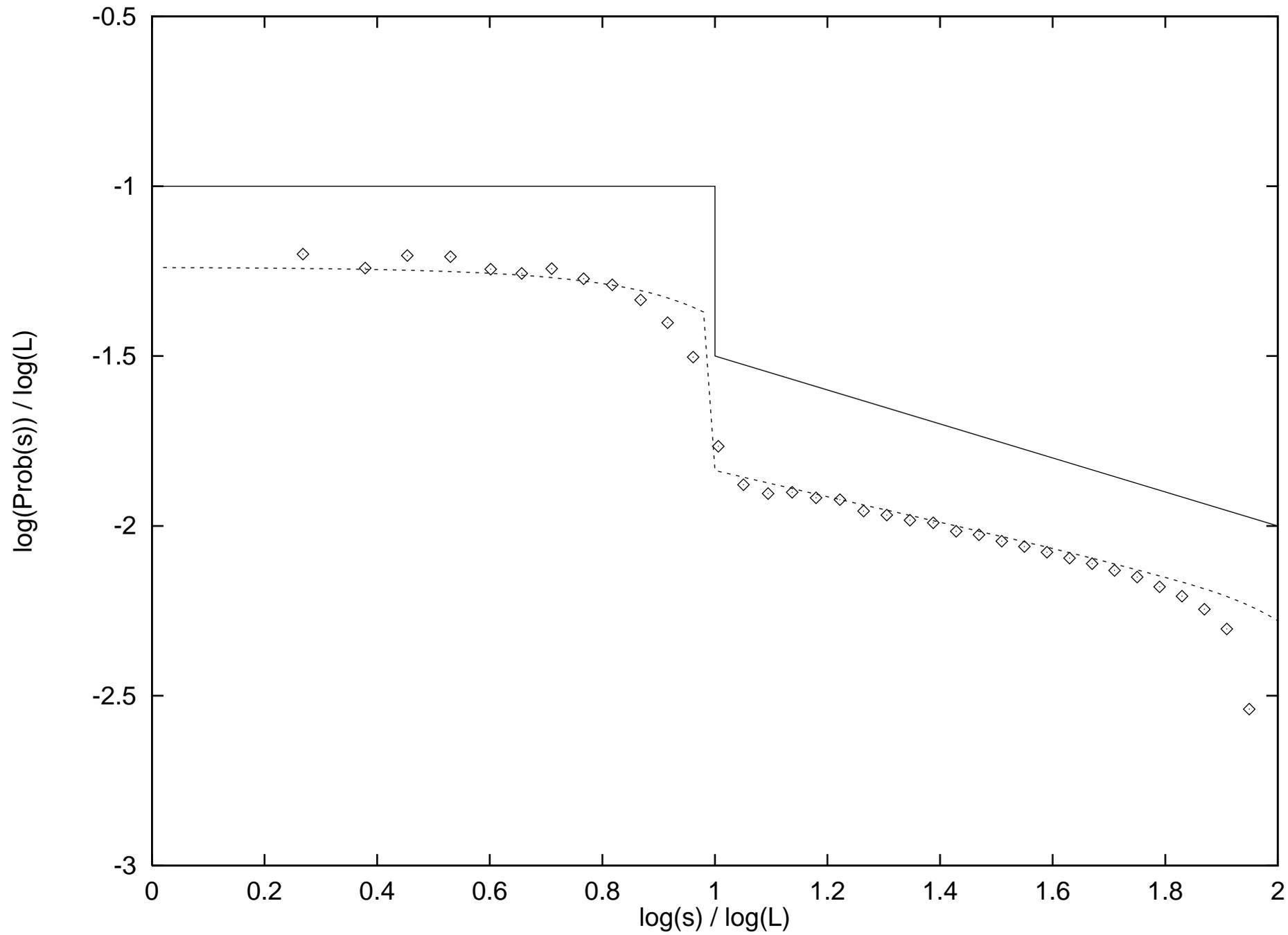
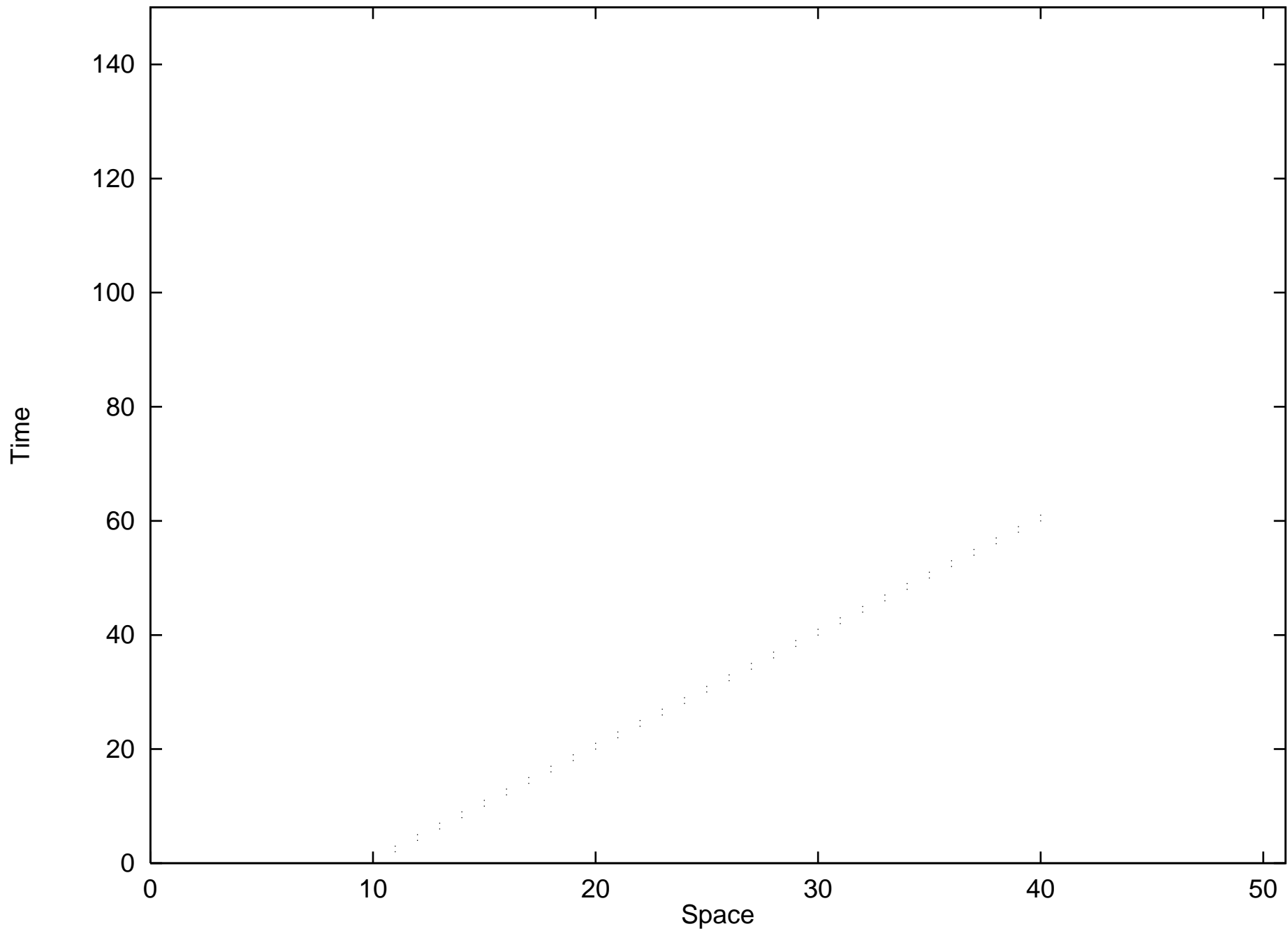
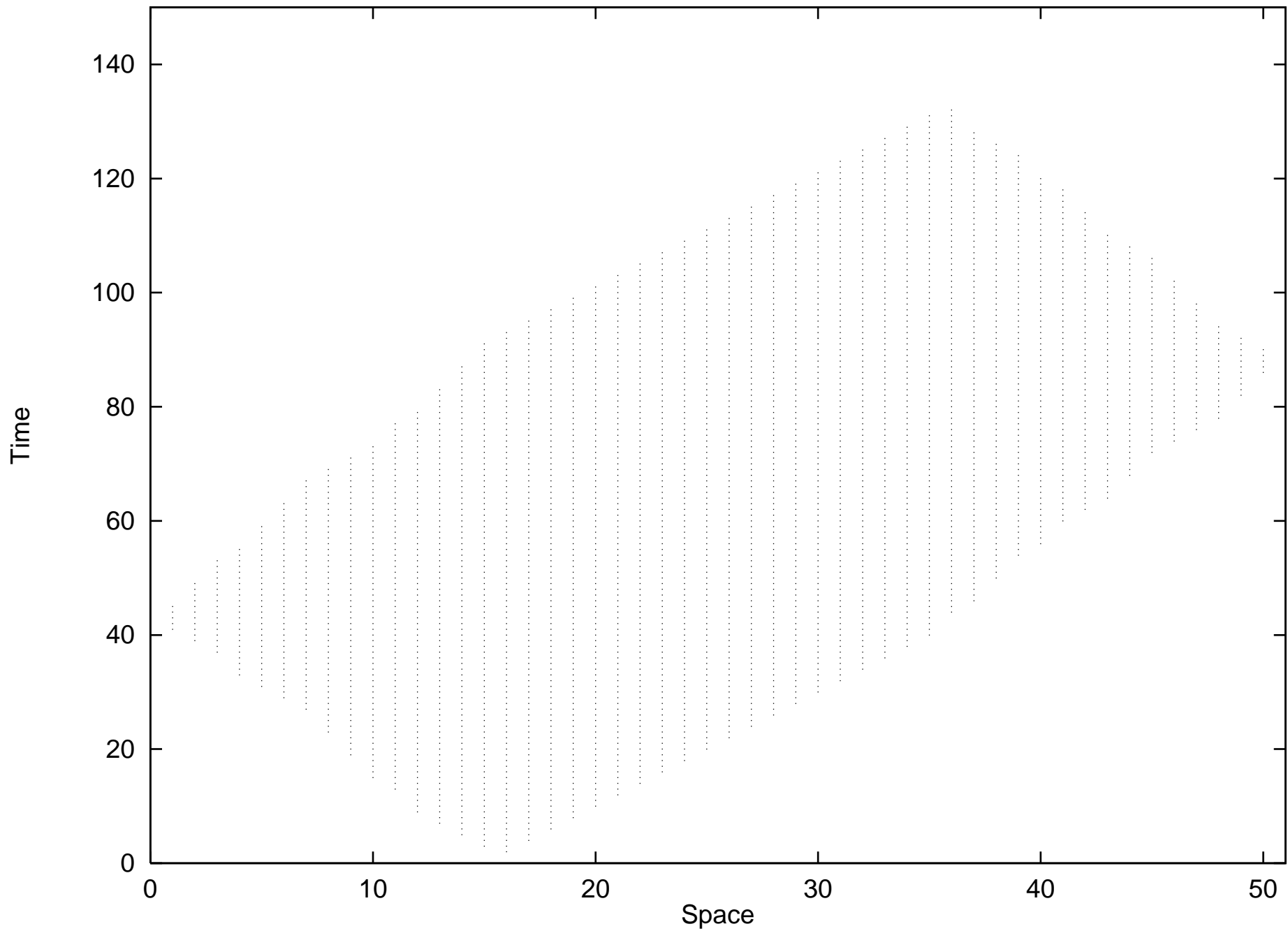
